\documentclass[useAMS,usegraphicx,usenatbib]{mn2e}

\newcommand{\nn}{\nonumber}
\def\lesssim{\buildrel < \over {_{\sim}}}
\def\gtrsim{\buildrel > \over {_{\sim}}}

\title[The role of ionization in the shock acceleration theory]{The role
of ionization in the shock acceleration theory}
\author[G. Morlino]{
Giovanni Morlino$^{1}$\thanks{E-mail: morlino@arcetri.astro.it} \\
$^{1}$INAF-Osservatorio Astrofisico di Arcetri, Largo E. Fermi, 5,
50125 Firenze, Italy}

\begin{document}

\date{Accepted -----. Received -----}

%\pagerange{\pageref{firstpage}--\pageref{lastpage}} \pubyear{2008}

\maketitle

\label{firstpage}

\begin{abstract}
We study the acceleration of heavy nuclei at SNR shocks taking into account the
process of ionization. In the interstellar medium atoms heavier then hydrogen
which start the diffusive shock acceleration (DSA) are never fully ionized at
the moment of injection. We will show that electrons in the atomic shells are
stripped during the acceleration process, when the atoms already move
relativistically. For typical environment around SNRs the dominant ionization
process is the photo-ionization due to the background Galactic radiation. The
ionization has two interesting consequences. First, because the total 
photo-ionization time is comparable to the beginning of the Sedov-Taylor phase,
the maximum energy which ions can achieve is smaller than the standard result 
of the DSA, which predict $E_{\max}\propto Z_N$. As a consequence the structure
of the CR spectrum in the {\it knee} region can be affected. The second
consequence is that electrons are stripped from atoms when they already
move relativistically hence they can start the DSA without any pre-acceleration
mechanism. We use the linear quasi-stationary approach to compute the spectrum
of ions and electrons accelerated after being stripped. We show that the number
of these secondary  electrons is enough to account for the synchrotron radiation
observed from young SNRs, if the amplification of the magnetic field occurs.
\end{abstract}


\section{Introduction}

The bulk of Galactic cosmic rays (CRs) is largely thought to be accelerated at
the shock waves associated with supernova remnants (SNRs) through the mechanism
of diffusive shock acceleration (DSA). A key feature of this mechanism is that
the acceleration rate is proportional to the particle charge. This property is
especially appealing if one tries to explain the structure of the knee in the CR
spectrum: if we assume that the maximum energy of accelerated particles scales
with the charge of the particles involved, a knee arises naturally as a
superposition of spectra of chemicals with different nuclear charges $Z_N e$
\cite[]{horandel03}. This result is based on the assumption that nuclei are
completely ionized during the acceleration. On the other hand when ions are
injected into the acceleration process they are unlikely to be fully stripped,
especially if of high nuclear charge. 

The atoms relevant for the injection are those present in the circumstellar
medium where the forward shock propagates. The temperature of this plasma
varies from $10^4$ K if the SNR expands into the regular ISM, up to $10^6$ K if
the expansion occurs into the bubble created by the progenitor's wind. If $T
\sim 10^4$ K even hydrogen is not fully ionized, as demonstrated by the
presence of Balmer lines associated with shocks in some young SNRs
\cite[]{chevalier80,sollerman,Heng09}. For $T \sim 10^6$ only atoms up to 
$Z_N= 5$ can be completely ionized \cite[]{porquet}.

The typical assumption made in the literature is that atoms lose all electrons
in the atomic orbitals soon after the beginning of the acceleration process,
namely that the ionization time needed to strip all the electrons is much
smaller than the acceleration time. In spite of this assumption in \cite{mor09}
we showed that, for a typical SNR shock, the ionization time is comparable with
the acceleration time, hence electrons are stripped when ions already move
relativistically. This fact has two important consequences: 1) the maximum
energy of ions can be reduced with respect to the standard prediction of DSA and
2) the ejected electrons can easily start the acceleration process because they
already move relativistically.

The possibility that ionization can provide a source of relativistic electrons
is especially relevant because the question of how electrons are injected into
the DSA is still an unsolved issue. DSA applies only for particles with a Larmor
radius larger than the typical shock thickness, which is of the order of Larmor
radius of the shocked downstream thermal ions. The injection condition can be
easily fulfilled for supra-thermal protons which reside in the highest energy
tail of the Maxwellian distribution. The injection of heavier ions is, in
principle, even easier than that of protons in that their Larmor radius is
larger assuming they have the same proton temperature (which could be not the
case).
On the other hand the same argument tells that electrons cannot be injected
from the thermal bath because, even if we assume equilibration between electrons
and protons, the Larmor radius of electrons is a factor $(m_e/m_p)^{1/2}$
smaller than that of protons. Only electrons which are already relativistic can
cross the shock and start the DSA.

Most proposed solutions to the electron injection problem involve some
kind of pre acceleration mechanism driven by plasma instabilities able to
accelerate thermal electrons up to mildly relativistic energy. For instance
\cite{galeev84} showed that magnetosonic turbulence excited by a beam of
reflected ions ahead of the shock can accelerate electrons along the magnetic
field lines thanks to their Cerenkov interaction with excited waves.
Other studies predict that electrons can be effectively preaccelerated by
self-generated whistler waves \cite[]{levinson92,levinson}. More recently
\cite{amano09} showed that efficient electron acceleration can happen in
perpendicular shocks due to ``shock surfing'' of electrons on electrostatic
waves excited by Buneman instability at the leading edge of the shock foot. A
different mechanism proposed by \cite{riquelme10}, can take place in
quasi-perpendicular shocks and is driven by the presence of oblique whistler
waves excited by the returning ions in the shock foot. The study of these
electromagnetic pre-acceleration mechanisms can be only performed using
numerical techniques which showed that the fraction of injected electrons
strongly depend on the values of initial conditions (like magnetic field
strength and orientation) and are difficult to apply to realistic cases.

In this paper we investigate in detail the process of ionization applied to the
acceleration of heavy ions, presenting the full steady-state solution for shock
acceleration in the test particle approach.

The paper is organized as follows: in \S\ref{sec:time} we compare the
acceleration time of ions with the ionization time due both to photo-ionization
and to Coulomb scattering showing that the former dominates on the latter. In
\S\ref{sec:E_max} we use the linear acceleration theory to compute the maximum
energy achieved by different chemical specie when the ionization process is
taken into account. In \S\ref{sec:spectra} we compute the distribution
function of both ions and electrons in the framework of linear acceleration
theory including the term due to ionization. We conclude in
\S\ref{sec:conclusion}.

\section{Ionization vs. acceleration time} \label{sec:time}

In this section we show that the ionization of different chemical species
occurs during the acceleration process on a timescale which is comparable with
the acceleration time needed to achieve relativistic energies. We also show
that the ionization due to Coulomb collision is generally negligible compared
with the photo-ionization due to the interstellar radiation field (ISRF).

Let us consider atoms of a single chemical specie $N$ with nuclear charge $Z_N$
and mass $m_N= A m_p$, which start the DSA with initial charge $Z<Z_N$ and
momentum $p_{\rm inj}$. We want to compute which is the ionization time needed
to lose one electron, changing the net charge from $Z$ to $Z+1$ and which is
the momentum $p$ that ions reach when ionization occurs.

For simplicity we compute the acceleration time in the framework of linear shock
acceleration theory for plane shock geometry: i.e. we assume that during the
ionization time needed to strip one single electron, the shock structure does
not change. If a particle with momentum $p$ diffuses with a diffusion
coefficient $D(p)$, the well know expression for the acceleration time is: 
\begin{equation}
 \tau_{acc}(p_{\rm inj}, p)= \int_{p_{\rm inj}}^{p}\frac{3}{u_1-u_2} \left(
\frac{D_1(p)}{u_1} +
 \frac{D_2(p)}{u_2} \right) \frac{dp}{p} \,,
 \label{eq:t_acc1}
\end{equation}
where $u$ is the plasma speed in the shock rest frame, and the subscript 1
(2) refers to the upstream (downstream) quantities (note that $u_{\rm shock}
\equiv u_1$). The downstream plasma speed is related to the upstream one
through the compression factor $r$, i.e. $u_2 = u_1/r$. We limit our
considerations to strong shocks, which have compression factor $r=4$, and we
assume Bohm diffusion coefficient, i.e. $D_B=r_L \beta c/3$, where $\beta c$ is
the particle speed and $r_L= pc/Z eB$ is the Larmor radius. The turbulent
magnetic field responsible for the particle diffusion is assumed to be
compressed downstream according to $B_2= r\,B_1$. Even if this relation applies
only for the magnetic component parallel to the shock plane, such assumption
does not affect strongly our results. 
It is useful to define the instantaneous acceleration time as $t_{acc} \equiv
d\tau/(dp/p)$. Using all previous assumptions $t_{acc}$ can be expressed as
follows:
\begin{equation}
 t_{acc}(p)= 0.85\, \frac{\beta\, p}{m_N c} \, B_{\mu G}^{-1} \, u_8^{-2} 
   \left( \frac{A}{Z} \right) \,{\rm yr} \,.
 \label{eq:t_acc2}
\end{equation}
Here the upstream magnetic field is expressed in $\mu$G and the shock speed is
$u_1= u_8\, 10^8 {\rm cm/s}$. Notice that, when $B$ and $u$ are assumed
constant, $\tau_{acc}$ reduces to $t_{acc}$ for $p_{\rm inj} \ll p$ and $D
\propto p$. 
To compute the energy reached by particles when the ionization event occurs,
Eq.~(\ref{eq:t_acc2}) has to be compared with the ionization time scale.
Ionization can occur either via Coulomb collisions with thermal particles or via
photo-ionization with background photons. Whether the former process dominates
on the latter depends on the ISM number density compared to the ionizing photon
density. In the following we show that photo-ionization dominates when SNRs are
young and expand in a typical ISM.

As soon as the ions start the DSA, the photo-ionization can occur only when
the energy of background photons, $\epsilon'$, as seen in the ion rest frame,
is larger than the ionization energy $I$. Atoms moving relativistically
with a Lorentz factor $\gamma$ see a distribution of photons peaked in the
forward direction of motion, with a mean photon energy $\epsilon'= \gamma
\epsilon$.  The photo-ionization cross section can be estimated using the 
simplest approximation for the $K$-shell cross section of hydrogen-like atoms
with effective nuclear charge $Z$, i.e. \cite[]{heitler}:
\begin{equation}
 \sigma_{\rm ph}(\epsilon') = 64\,\alpha^{-3}\sigma_T Z^{-2}
  \left( I_{N,Z}/\epsilon'\right)^{7/2} \,
 \label{eq:Photoion}
\end{equation}
where $\sigma_T$ is the Thompson cross section and $\alpha$ is the fine
structure constant and $I_{N,Z}$ is the ionization energy threshold for the
ground state of the chemical specie $N$ with $Z_N-Z$ electrons. The numerical
values of $I_{N,Z}$ can be found in the literature (see e.g. \cite{allen73}
for elements up to $Z_N=30$). To get the full photo-ionization time we need to
integrate over the total photon energy spectrum, i.e.:
\begin{equation} \label{eq:Ph_time}
 \tau_{\rm ph}^{-1}(\gamma)= \int d\epsilon\, 
 \frac{dn_{\rm ph}(\epsilon)}{d\epsilon} \,c\,
     \sigma_{\rm ph}(\gamma \epsilon)  \,,
\end{equation}
where $dn_{ph}/d\epsilon$ is the photon spectrum as seen in the plasma rest
frame. Because the photo-ionization cross section decreases rapidly with
increasing photon energy, for a fixed ion speed the relevant ionizing photons
are only those with energy close to the threshold, i.e. $\epsilon \simeq
I_{N,Z}/\gamma$, measured in the plasma frame.
The corresponding numerical value is:
\begin{equation}
 \tau_{\rm ph}(\gamma) \simeq 0.01\, Z^2 \, \left( n_{\rm ph}(I_{N,Z}/\gamma)
 /{\rm cm^{-3}} \right)^{-1} \, {\rm yr} \,.
 \label{eq:Ph_time2}
\end{equation}
We assume that the maximum possible acceleration time is equal to the
Sedov-Taylor time, $t_{ST}$, corresponding to the end of the free expansion
phase. Comparing $t_{ST}$ with Eq.~(\ref{eq:Ph_time2}), we see that the photons
which can be relevant for ionization are only those with a number density
$n_{\rm ph}> 10^{-5} Z^2 (t_{ST}/10^3 \rm yr)^{-1} \,cm^{-3}$. For this reason
we can neglect the high energy radiation coming from the remnant itself because
the typical number density of the x-ray photons is less than $10^{-7}$
ph/cm$^{-3}$.

An important consequence of the photo-ionization is that electrons are stripped
from the atoms with a kinetic energy $\ll m_e c^2$ as measured in the atom rest
frame. Hence, if the parent atoms move relativistically in the plasma rest
frame, electrons move with the same Lorentz factor and the same direction
of the parent atoms. This property will be used in \S \ref{sec:spectra} to
simplify the calculation of electrons spectrum.

Now we consider the ionization of accelerated ions due to Coulomb collisions
with protons and electrons of the thermal plasma. We neglect the thermal energy
of those particles and we consider the process in the ion's rest frame.
The role of collisions is mainly relevant during the very initial stage of
acceleration, namely when the kinetic energy of incident particle, $E_{\rm
kin}$, is close to the ionization energy, $I$, as seen in the ion's rest frame.
This is a consequence of the shape of ionization cross section which has a peak
value for $E_{\rm kin} \sim I$ and rapidly decreases for larger energies. A
good upper limit for the peak value of the cross section is provided by the
classical (non relativistic) Thomson approach \cite[]{allen73}, which gives a
maximum for $E_{\rm kin}=2 I$ equal to:
\begin{equation}
 \sigma_{\rm coll}^{\rm max} = \pi a_0^2 N_e\, I_{\rm Ryd}^{-2} \,.
 \label{eq:Coulomb}
\end{equation}
where $a_0$ is the Bohr radius, $I_{\rm Ryd}$ is the ionization energy
expressed in Rydberg unity and $N_e$ is the number of electrons in the
atomic orbital considered. For larger energies the Thomson cross section
decreases like $E_{\rm kin}^{-2}$ while the asymptotic behavior predicted by
the quantum mechanics is $\sigma \sim \log(E_{\rm kin})/E_{\rm kin}$. 
This implies that the collisional ionization time has a minimum for
$\sigma = \sigma_{\rm coll}^{\rm max}$ and than increases. In order to get this
minimum value one must average the contribution of collisions in the upstream
and downstream region. Assuming equal density for thermal protons and electrons
$n_e=n_p \equiv n_1$, the result is \cite[]{mor09}:
\begin{equation}
 \tau_{\rm coll}^{\rm min} \approx \left[c \, \sigma_{\rm coll}^{\rm max}\, 
    n_1 (1+r)\right]^{-1} 
  = 0.0024 \, I_{\rm Ryd}^2 \left( \frac{n_1}{\rm cm^{-3}}\right)^{-1}
    \, {\rm yr} \,.
 \label{eq:Cion_time}
\end{equation}
Equation ~(\ref{eq:Cion_time}) is valid in the non relativistic regime. In this
regime the acceleration time increases only like $t_{\rm acc} \sim p \sim E_{\rm
kin}^{1/2}$. As a consequence the collisional ionization can occur only if the
acceleration time needed to have $E_{\rm kin}= 2 I$ is larger than $\tau_{\rm
coll}^{\rm min}$. Let us call $\gamma^*$ the Lorentz factor which corresponds to
this condition, namely $\gamma^*=1 + 2 I/m_e c^2$. Comparing
Eq.~(\ref{eq:Cion_time}) with Eq.~(\ref{eq:t_acc2}), the condition 
$t_{\rm acc}(\gamma^*) = \tau_{\rm coll}^{\rm min}$ provides a lower limit for
the upstream density, i.e.:
\begin{equation}
 n_{1,\rm min} = 27 \, I_{\rm Ryd} B_{\mu G} u_8^2 \left(Z/Z_N \right) 
   {\rm cm^{-3}} \,.
 \label{eq:n1_max}
\end{equation}
Hence for $n_1 < n_{1,\rm min}$ ions are accelerated up to relativistic energies
before being ionized by collisions. Notice that we neglected collisions
with atoms heavier than hydrogen: such contribution could reduce the ionization
time by less than a factor 2, hence are main conclusions remain valid.

Now, in order to understand the relevance of collisions in the relativistic
regime we can use the asymptotic Bethe cross section valid in the limit $E_{\rm
kin} \gg m_e c^2$, \cite[]{Bethe30}. The relativistic formulation of this cross
section for the total ionization is often written as \cite[]{kim00}:
\begin{equation}
 \sigma_{\rm RBehte}(\gamma)= \frac{4 \pi a_0^2 \alpha^2}{\beta^2 } 
 \left[ M^2 \left( \ln(\gamma^2
 \beta^2)-\beta^2 \right) + C_R \right] \,.
 \label{eq:Bethe}
\end{equation}
The two constant $M^2$ and $C_R$ are related to the atomic form factors of the
target and are independent of the incident particle energy. Their exact
numerical values are very difficult to compute with some noticeable exception
like H-like atoms. In principle they can also be inferred from photo-ionization
experiments but in practice they are known only for a bunch of targets.
Nevertheless from theory we know that $M^2\sim N_e/I_{\rm Ryd}$, while $C_R
\sim 10 M^2$ \cite[]{inokuti, kim00}. Using these estimates in the ultra
relativistic limit we have $\sigma_{\rm RBethe}/\sigma_{\rm coll}^{\max}
\lesssim 10^{-3}$. This result implies that Coulomb collisions are much less
important in the relativistic regime with respect to the non relativistic one.

\begin{figure}
\begin{center}
{\includegraphics[angle=0,width=0.9\linewidth]{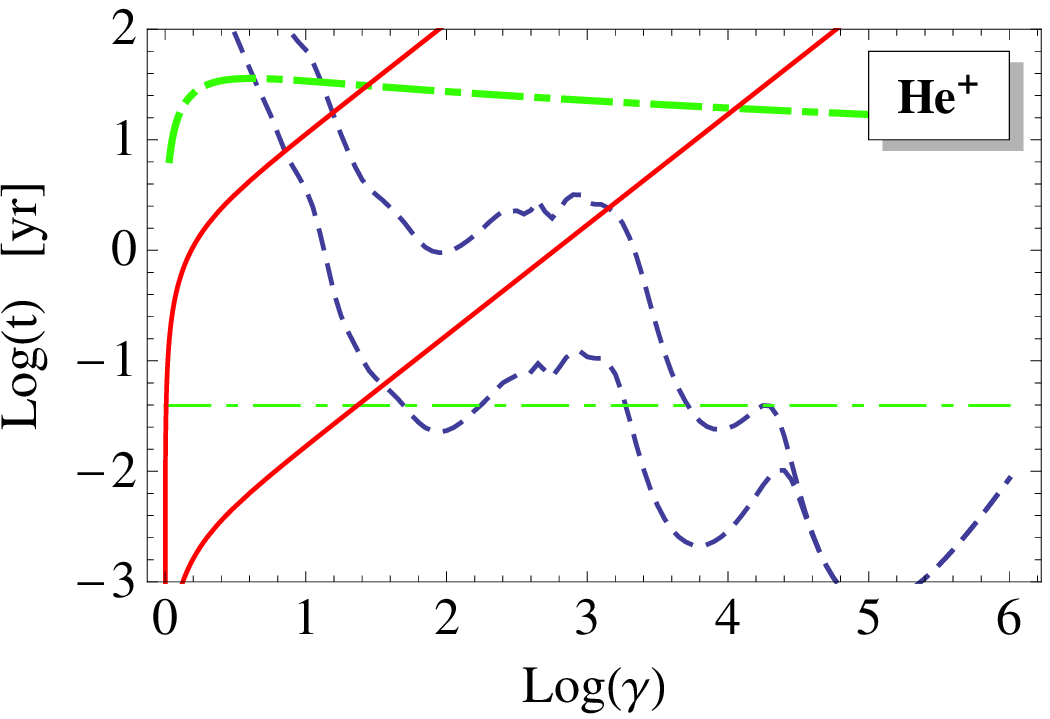}}
{\includegraphics[angle=0,width=0.9\linewidth]{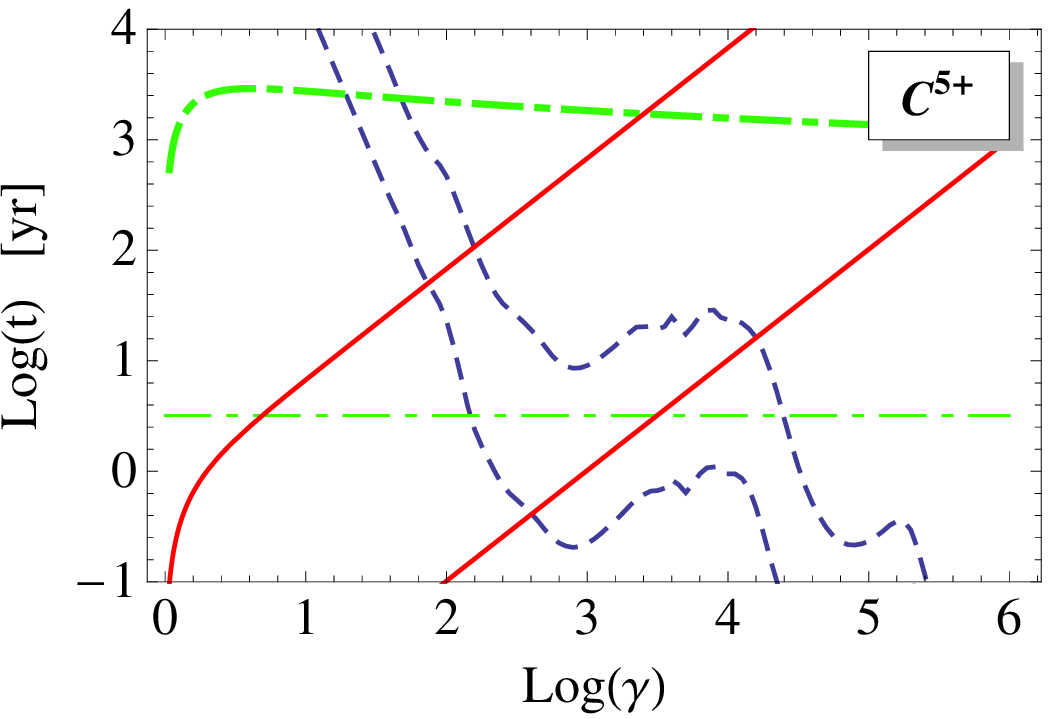}}
{\includegraphics[angle=0,width=0.9\linewidth]{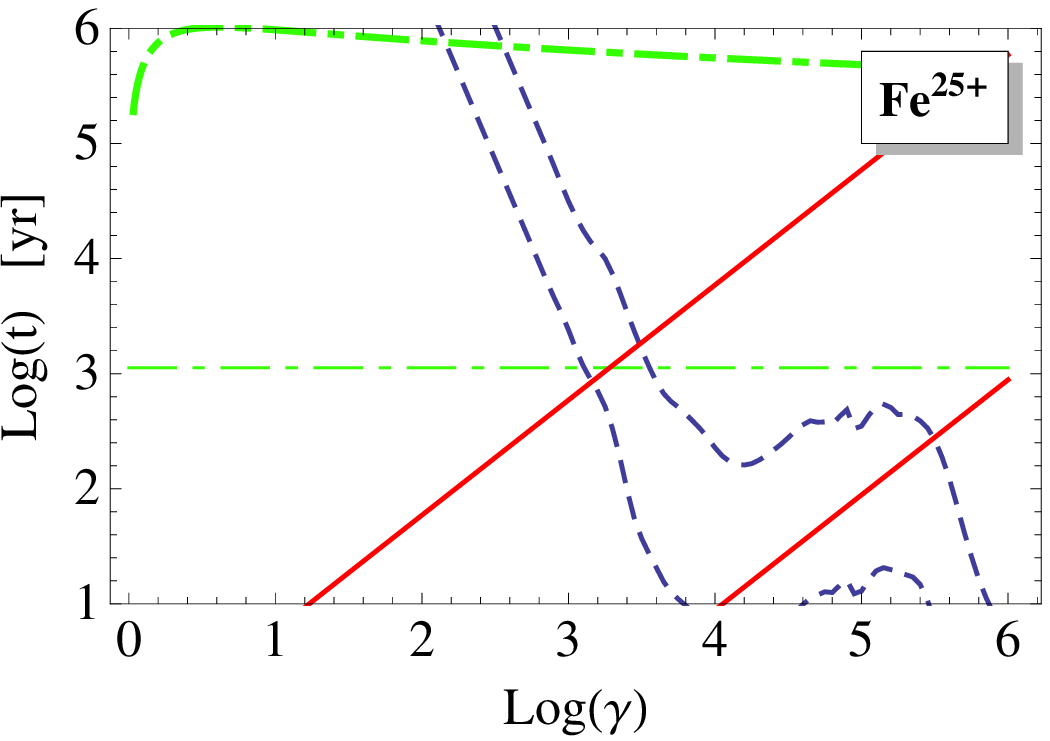}}
\caption{ Comparison between acceleration time (solid lines),
photo-ionization time (dashed) and collisional ionization time
(dot-dashed) for three hydrogen-like ions: He$^{+}$, C$^{5+}$ and
Fe$^{25+}$ (from top to bottom). Acceleration and photo-ionization times are
shown with two curves representing two different set of parameters, as explained
in the text. The horizontal thin dot-dashed line represent the lower limit for
the collisional ionization time, Eq.~(\ref{eq:Coulomb}), while the thick
dot-dashed line is computed using the Bethe cross section with a plasma
density $n_1= 1 {\rm cm^{-3}}$.}
\label{fig:time}
\end{center}
\end{figure}

In order to compare the photo-ionization with the collisional ionization times,
in Fig.~\ref{fig:time} they have been plotted together with the acceleration
time for three different H-like ions: He$^{+}$, C$^{5+}$ and Fe$^{25+}$. The
characteristic time are plotted as functions of the ion's Lorentz factor.
The acceleration time $t_{\rm acc}$, plotted with solid lines, is shown for two
different choices of parameters: $u_1= 1000$ km/s ; $B_1= 3 \mu$G (upper line)
and for $u_1= 10^4$ km/s, $B_1= 20 \mu$G (lower line). The photo-ionization time
$\tau_{\rm ph}$ (dashed lines) is computed according to Eq.~(\ref{eq:Ph_time})
using the cosmic microwave background plus the Galactic ISRF as reported by
\cite{porter05}. The ISRF strongly changes going from the center towards the
periphery of the Galaxy, hence, the photo-ionization depends on the SNR
location. In order to evaluate this variation we show $\tau_{\rm ph}$ using the
ISRF in the Galactic Center (lower dashed line) and the one at 12 kpc far away
from the Galactic Center in the Galactic plane (upper dashed lines). Finally
with dot-dashed lines we plot the collisional ionization time for the upstream
density $n_1= 1 \rm cm^{-3}$: the thin horizontal line corresponds to the non
relativistic lower limit expressed in Eq.~(\ref{eq:Cion_time}), 
$\tau_{\rm coll}^{\rm min}$, while the thick line is calculated using the
relativistic Bethe cross section for H-like atoms.
In this case the values of the parameters used in the Bethe cross section are
$M^2=0.30/Z_N^4$ and $C_R= 4.30/Z_N^4$ \cite[]{Bethe30}. 
Fig.~\ref{fig:time} clearly shows that the collisional ionization can be
neglected for the case of young SNRs expanding into a medium with a number
density less than a few cm$^{-3}$. The photo-ionization is the dominant
process even for SNR located far away from the Galactic Center. This
consideration is strengthened in the case of core-collapse SNRs, which typically
expand into the bubble created by the progenitor wind, whose density is
usually assumed $\sim 10^{-2} \rm cm^{-3}$. The only phase where the collisional
ionization could dominate over photo-ionization is the very initial stage, when
the shock is expanding into the progenitor's wind, whose density is much higher.
This phase typically lasts few years and in this work it will be neglected. 

In Fig.~\ref{fig:time} the value of $\gamma$ where $t_{\rm acc}$ cross
$\tau_{\rm ph}$, identifies the Lorentz factor of the ejected electrons they
are for. Looking at the upper plot it is clear that even electrons coming from
the ionization of He$^+$ can, in principle, start the DSA, having a typical
Lorentz factor $\gtrsim 10$. In fact, as discussed in \cite{mor09} the
minimum Lorentz factor required for electrons to be injected is 
$\gamma_{\rm inj} \sim 3-30$ for typical parameters of SNR shocks.

\section{Maximum energy of Ions}  \label{sec:E_max}
The process of ionization can affect the maximum energy that nuclei achieve
during the acceleration, especially those with large nuclear charge. It is worth
stressing that the knowledge of the maximum energy is intimately connected with
two important aspects of the CR spectrum: the interpretation of the {\it knee}
structure and the transition region from Galactic to extragalactic component.
The {\it knee} is commonly interpreted as due to the superposition of the
spectra of all chemicals with different cutoff energies. Using the flux of
different components measured at low energies it has been shown that the knee
structure is well reproduced if one assumes that the maximum energy of each
specie $E_{\max,N}$ is proportional to the nuclear charge $Z_N$
\cite[]{horandel03}.
Nevertheless the superposition of subsequent cutoff seems to be confirmed by the
measurements of the spectrum of single components in the knee region: data
presented by the KASCADE experiments show that the maximum energy of He is $\sim
2$ times larger than that of the protons \cite[]{antoni05}.
From the theoretical point of view the relation $E_{\max,N} \propto Z_N$ is
clearly predicted by DSA if one assume that the diffusion coefficient is
rigidity-dependent. On the other hand this is correct only provided that during
the acceleration process nuclei are completely ionized otherwise the maximum
energy is proportional to the effective charge of the ions rather that to 
their nuclear charge.

Now, the maximum energy is thought to be achieved at the beginning of the
Sedov-Taylor phase ($t_{ST}$) \cite[]{BAC-pmax07}. For later times, $t>t_{ST}$,
the shock speed decreases faster than the diffusion velocity, hence particles at
the maximum energy can escape from the accelerator and the maximum energy
cannot increase further. It is worth noting that the process of escaping is not
completely understood in that it strongly depends on the turbulence produced by
the same escaping particles and on the damping mechanisms of the turbulence
itself \cite[]{caprioli10b}. Anyway here we assume that the maximum energy,
$E_{\max}$, is achieved at $t=t_{ST}$. If the total ionization time is
comparable, or even larger than $t_{ST}$, we do expect that $E_{\max} <
E_{\max}^0$, where we call $E_{\max}^0$ the maximum energy achieved by ions
which are completely ionized since the beginning of acceleration.

A consistent treatment of the ionization effects would require the use of
time-dependent calculation. However our aim is to get a first order
approximation of the effect of ionization. This can be achieved using the
quasi-stationary version of the linear acceleration theory.

In order to compute $E_{\max}$ the first piece of information we need is the
initial ionization level of ions. In the case of volatile elements, which exist
mainly in the gas-phase, the level of ionization is easy to estimate being only
a function of the plasma temperature: ions are ionized up to a level such
that the ionization energy is of the same order of the kinetic energy of thermal
electrons as seen in the ion's rest frame. On the other hand refractory elements
are largely locked into solid dust grains. This occurs both in the regular ISM
as well as in stellar winds. As claimed by \cite{ellison97} there are
evidences that elements condensed in dust grains are efficiently injected into
the DSA thanks to the fact that dust grains can be easily accelerated up to
energies where the grains start to be sputtered. A non negligible fraction of
the atoms expelled by sputtering are energetic enough to start the DSA. In this
scenario the ionization level of ejected atoms is very difficult to predict in
that depends on the structure of the grains. However a reasonable assumption
could be that ions preserve all the electrons in the inner shells which are
not shared in the orbitals of the crystal structure of the grains. Moreover, as
noted by \cite{ellison97}, if the ejected ion is highly ionized its electron
affinity is strong and electron-exchange with the atoms of thermal plasma could
reduce the level of ionization up to the equilibrium one.
For these reasons, in the following calculation, we neglect the complication
arising from the dust sputtering process and we assume that the level of
ionization is only determined by the plasma temperature.

Let us consider a single ion injected with momentum $p_0$ and total charge
$Z_1$. The ion undergoes acceleration at a constant rate $\propto Z_1$ during a
time equal to the ionization time needed to lose one electron, $\tau_{\rm
ph,1}$, when it achieves the momentum $p_1$. Notice that we include only the
photo-ionization because, as we saw in the previous paragraph, for typical
density of the external medium around the SNR ($n_0 \lesssim 1 {\rm cm^{-3}}$),
the Coulomb collisions can be neglected. In order to compute simultaneously
$\tau_{\rm ph,1}$ and $p_1$ we have to equate the acceleration time with the
ionization time:
\begin{equation}
 \tau_{\rm ph,1}(p_1/m_Nc) = \tau_{acc}(p_0,p_1) 
  = t_{\rm acc}(p_1) - t_{\rm acc}(p_0) \,.
 \label{eq:p2}
\end{equation}
The last equality holds because we use Bohm diffusion and because we assume that
the shock velocity and the magnetic field both remain constant during $\tau_{\rm
ph,1}$. Moreover we saw that the photo-ionization occurs when ions already move
relativistically, hence using Eq.~(\ref{eq:t_acc2}) we set $\beta=1$.
Eq.~(\ref{eq:p2}) gives
\begin{equation}
 \frac{p_1}{m_p c}= \frac{p_0}{m_p c} + \frac{Z_1 B_1 u_1^2}{1.7 \, Z_N} \,
                    \tau_{\rm ph,1}\left(p_1/m_N c\right) \,,
 \label{eq:p2_sol}
\end{equation}
where the subscript $1$ label the quantities during the time $\tau_{\rm ph,1}$.
In Eq.~(\ref{eq:p2_sol}) $B$ and $u$ are expressed in units of $\mu G$ and
$u_8$ while $\tau_{\rm ph,1}$ is expressed in yr. Once the background
photon distribution is known Eq.~(\ref{eq:p2_sol}) can be solved numerically in
order to get $p_1$ and $\tau_{\rm ph,1}$. 
After the first ionization the acceleration proceeds at a rate proportional to
$Z_2 \equiv Z_1+1$ during a time $\tau_{\rm ph,2}$, which is the time needed to
lose the second electron. Applying Eq.~(\ref{eq:p2_sol}) repeatedly for all
subsequent ionization steps, we get the momentum when the ionization is
complete. If the total ionization time, $\tau_{\rm ion}^{\rm tot}$, is
smaller than the Sedov time, in order to get the maximum momentum we need to add
the further acceleration during the time $(t_{ST} - \tau_{\rm ion}^{\rm tot})$.
The final expression for the maximum momentum can be written as follows:
\begin{equation}
 \frac{p_{\max}}{m_pc}=  \sum_{k=1}^{Z_N-Z_0} 
      \frac{Z_k B_k u_k^2 \tau_{\rm ph,k}}{1.7\, Z_N}
    + \theta\left(t_{ST}-\tau_{\rm ion}^{\rm tot}\right)
      \sum_{i=1}^{M} \frac{B_i u_i^2}{1.7\, M}\,,
 \label{eq:p2_sol2}
\end{equation}
where we neglected the contribution of $p_0$. The last term has been written
as a sum over $M$ time-steps in order to handle the case where magnetic field
and shock speed change with time.

Now we can quantify the effect of ionization in the determination of $p_{\max}$.
The simplest approximation we can do is to assume $u_{sh}$ and $B$ constant
during the free expansion phase, i.e. up to $t=t_{ST}$.
In order to estimate $t_{ST}$ we consider two different situations which can
represent a typical type I/a and a core-collapse supernovae. In the first case
the supernova explodes in the regular ISM with typical density and temperature
$n_1= 1$ cm$^{-3}$ and $T_0= 10^4$ K. On the other hand SNRs generated by very
massive stars expand into medium with higher temperature. This could be
either the bubble generated by the progenitor's wind or a so called
\textit{super-bubble}, a region where an elevate rate of SN explosions produces
a diluted and hot gas \cite[]{higdon}. In both cases typical values for the
density and temperature inside the bubble are $n_1= 10^{-2}$ cm$^{-3}$ and $T_0=
10^6$ K.  For both type I/a and core-collapse supernovae we assume the same
value for the explosion energy $E_{SN}= 10^{51}$ erg, and mass ejecta $M_{ej}=
1.4 M_{\odot}$. The resulting Sedov time is $t_{ST}= 470$ yr and $t_{ST}= 2185$
yr, respectively. The average shock speed during the free expansion phase is
$u_{sh}\simeq 6000$ km/s in both cases.

It is worth stressing that if one would consider a more realistic scenario for
core-collapse SNe it is important to take into account the type of the
progenitor and its wind, in order to provide a better estimate of the ejected
mass and of the bubble's dimension and density. These parameters strongly
affect the value of $t_{ST}$, which determines how effective is the ionization
in reducing the maximum energy: indeed for $t_{ST} \gg 10^3$ yr the role of
ionization becomes negligible, as we see below.

The last parameter we need is the magnetic field strength. We stress that our
aim is only to estimate the effect of ionization in the scenario where the SNRs
indeed produce the observed CRs spectrum and the {\it knee} is interpreted as
the superposition of a rigidity-dependent cutoff of different species. For this
reason we assume that both scenarios are able to accelerate protons up to the
{\it knee} energy, i.e. $E_{\rm knee}= 3\cdot 10^{15}$ eV. This condition gives
$B_1= 160 \mu G$ and $35 \mu G$ for type I/a and core-collapse cases,
respectively. We note that the chosen values of magnetic field strength are
consistent with those predicted by the CR-induced magnetic filed amplification.

\begin{figure}
\begin{center}
{\includegraphics[angle=0,width=1.0\linewidth]{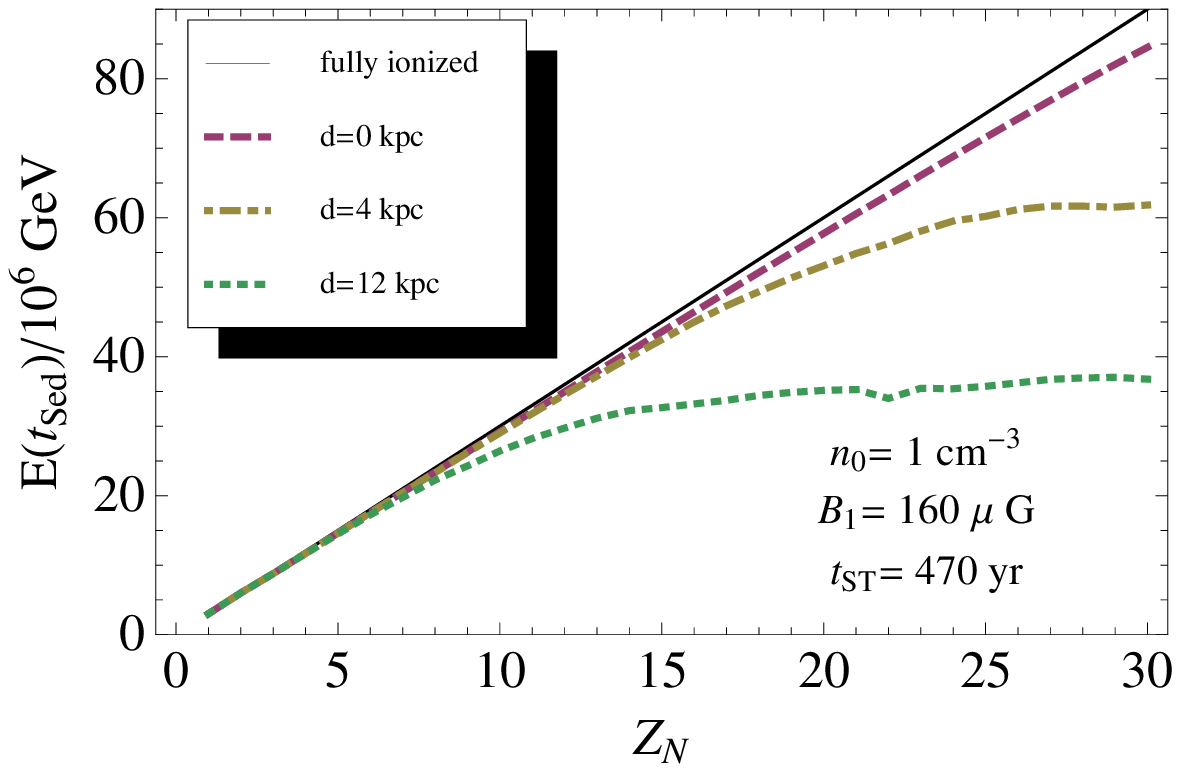}}
{\includegraphics[angle=0,width=1.0\linewidth]{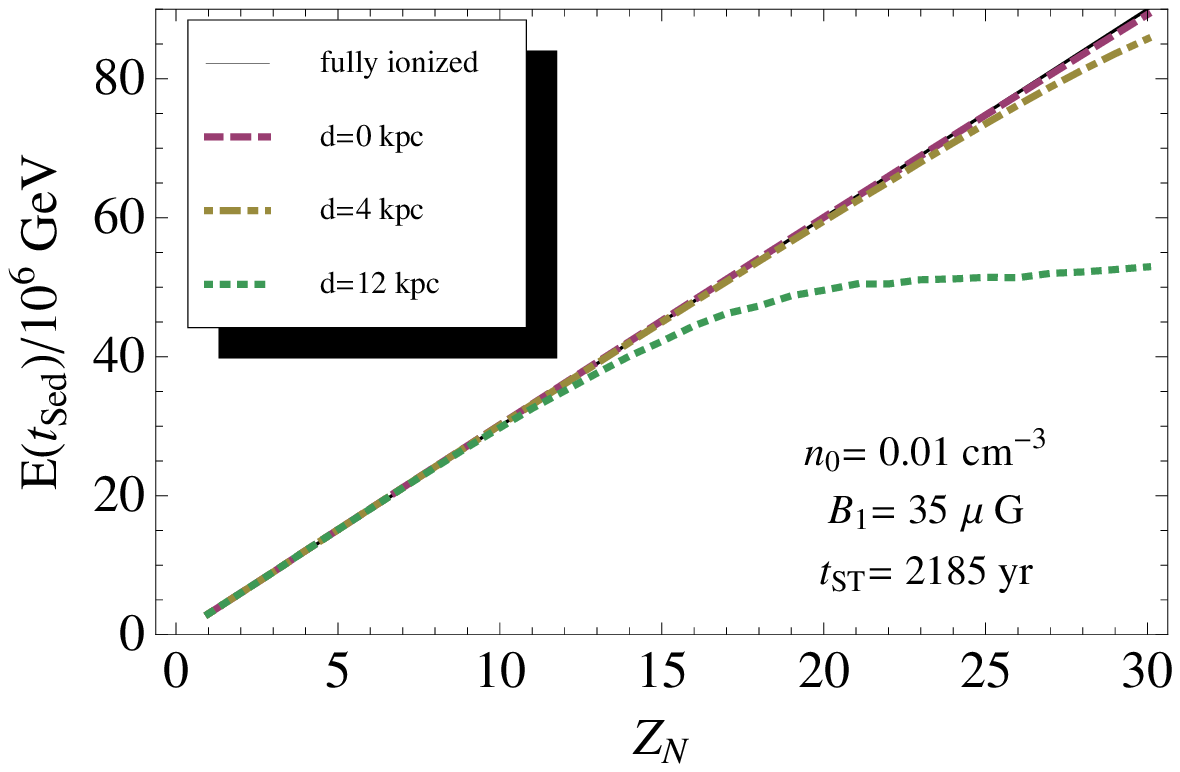}}
\caption{Maximum energy achieved by chemical specie with nuclear charge $Z_N$
(ranging from H to Zn) at the beginning of the Sedov phase. The upper
panel shows the case of type I/a SNR, expanding into a medium with density
$\rho_0=1\, {\rm cm^{-3}}$ while the lower panel is for core-collapse SNR
expanding into a diluted bubble ($\rho_0= 0.01 {\rm cm^{-3}}$). The thin solid
line shows the maximum energy achieved by atoms which are fully ionized since
the beginning of the acceleration, while the other curves are computed including
the photo-ionization due to ISRF for SNR located at three different distance $d$
from the Galactic Center, as specified in the caption.}
\label{fig:EmaxI}
\end{center}
\end{figure}

In Fig.~(\ref{fig:EmaxI}) we plot the maximum energy achieved at the beginning
of the Sedov-Taylor phase by different chemical species, from H up to Zn
$(Z_N=30)$. The two panels show the case of type I/a (upper) and core-collapse
SNR (lower) previously described. Each panel contains four lines: thin solid
lines are the maximum energy achieved by ions which start the acceleration
completely stripped, $E_{\max}^0$, while the remaining lines show $E_{\max}$
computed according to Eq.~(\ref{eq:p2_sol2}) for three different locations of
the SNR in the Galactic plane: at the center of the Galaxy and at 4 and 12 kpc
away from the Galactic Center. Looking at the upper panel we see that the
maximum energy achieved by different nuclei in type I/a SNRs does not increase
linearly with $Z_N$, instead it reaches a plateau for $Z_N \gtrsim 25$ at a
distance $d=4$ kpc and for $Z_N \gtrsim 15$ at $d= 12$ kpc. Only for SNRs
located in the Galactic bulge the reduction of the maximum energy is negligible,
at least for elements up to $Z_N= 30$. The effect of ionization is much less
relevant for core collapse SNRs: only those remnants located at a distance of 12
kpc show a noticeable reduction of $E_{\max}$.

\begin{figure}
\begin{center}
{\includegraphics[angle=0,width=1.0\linewidth]{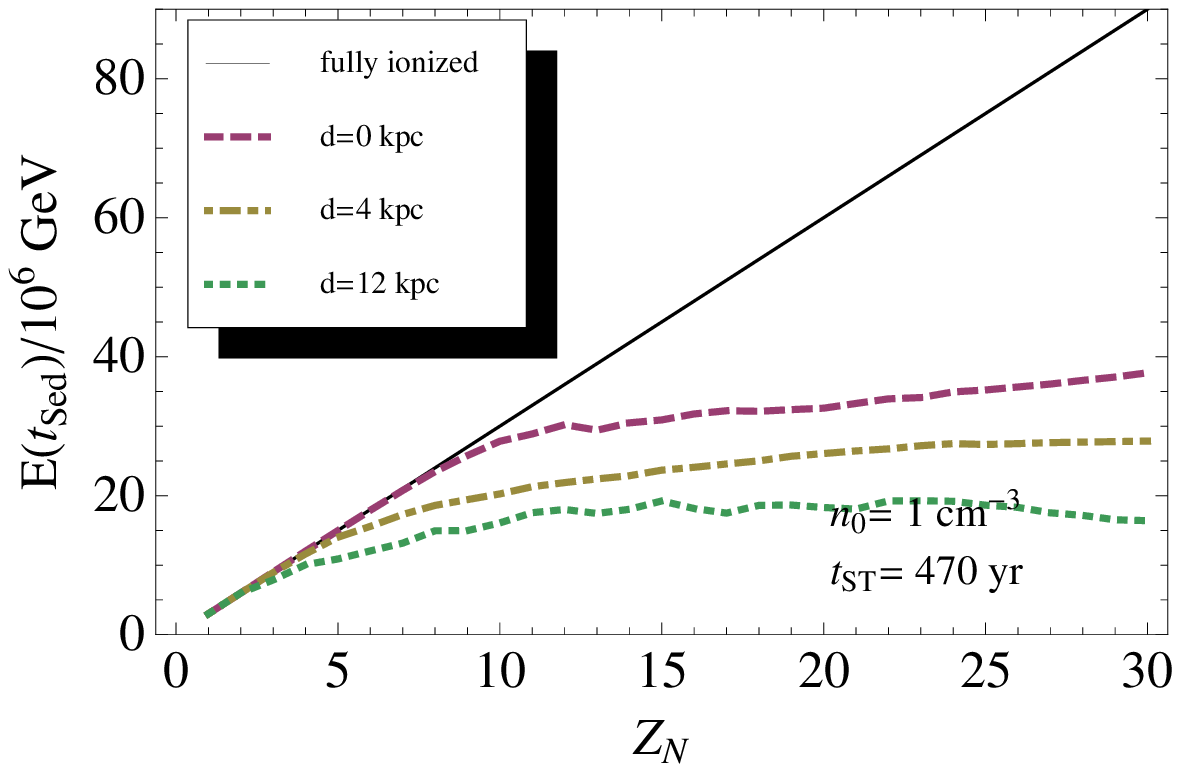}}
{\includegraphics[angle=0,width=1.0\linewidth]{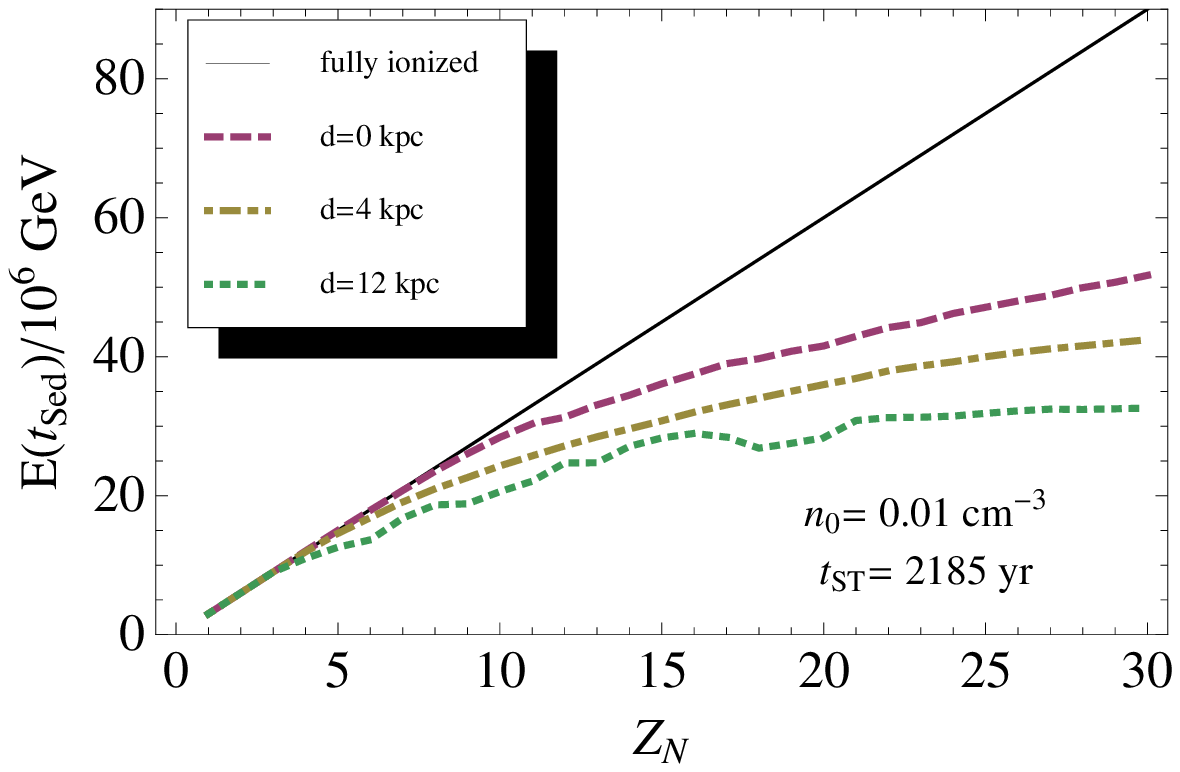}}
\caption{The same as in Fig.~\ref{fig:EmaxI} but for a scenario where both the
shock speed and the magnetic field strength evolve with time, as explained in
the text.}
\label{fig:EmaxII}
\end{center}
\end{figure}

\begin{table}
\begin{center}
\begin{tabular}{ccccc}
\hline
  & $t_{ST}$(yr) & $d$ (kpc) & $\tau_{\rm ion}^{\rm tot}$(yr) &
    $E_{\max}/E_{\max}^0$ \\
\hline
\hline
type I/a& 470  & 0    &   76 (137)    & 0.98 (0.52)   \\
        &  ''  & 4    &  356 (385)    & 0.80 (0.39)   \\
        &  ''  & 12   & 1446 (2932)   & 0.48 (0.24)   \\
\hline
core-      & 2185 & 0    &  113 (142)    & 0.99 (0.58)   \\
collapsed  &  ''  & 4    &  264 (374)    & 0.97 (0.48)   \\
           &  ''  & 12   & 2860 (2780)   & 0.65 (0.35)   \\
\hline
\end{tabular}
\caption{Total photo-ionization time and normalized maximum energy achieved by
iron nuclei for the two idealized cases of type I/a and core-collapsed SNRs,
which are located at a distance $d= 0, 4$ and 12 kpc from the Galactic Center.
For comparison also the Sedov time is listed. The bare numbers represent the
case where the shock speed and the magnetic field are both constant during the
free expansion phase (as in Fig.~\ref{fig:EmaxI}), while the numbers in round
brackets are computed assuming $u_{sh}$ and $B_1$ which evolve in time
according to the model explained in the text (see also Fig.~\ref{fig:EmaxII}).}
\label{tab:1}
\end{center}
\end{table}

The numerical results for iron nuclei corresponding to all cases shown in
Fig.~\ref{fig:EmaxI} are summarized in Table~\ref{tab:1}. Here we report the
total ionization time and the ratio between the energy achieved at $t_{ST}$ and
the maximum theoretical energy achieved by bare nuclei, $E_{\max}^0$ (notice
that the numbers in brackets refer to Fig.~\ref{fig:EmaxII} as explained
below).

The results presented in Fig.~\ref{fig:EmaxI} are inferred using a simple
steady state approach. This approach could be too reductive and one can guess
how the effect due to the ionization process changes when the full evolution
history of the remnant is taken into account. Needless to say a correct
calculation of the maximum energy requires not only the treatment of the
evolution of the remnant but also the inclusion of nonlinear effects which
consistently describe the magnetic field amplification and the back reaction of
CRs onto the shock dynamics. A fully consistent treatment of the problem is
beyond the purpose of this work. However we would understand whether the
evolution and the nonlinearity can reduce or exacerbate the effect induced by
ionization. 
In order to reach this goal we approximate the continuum evolution into time
steps equal to the ionization times, $\tau_{\rm ph,k}$ and we assume that the
stationary approximation is valid for each time step. Under this approximation
Eq.~(\ref{eq:p2_sol2}) can be used to include the effect of evolution if we
make reasonable assumptions on how the shock speed and the magnetic field change
with time. We follow \cite{truelove99} for the description of the time evolution
of the remnant. Specifically we adopt the solution for a remnant characterized
by a power-law profile of the ejecta (in the velocity domain) with index $n=7$,
expanding into an homogeneous medium. For this specific case \cite{truelove99}
provide the following expression for the velocity of the forward shock (see
their Table~7):
\begin{equation} \label{eq:u1_t}
 u_1(t)/u_{ch}= 0.606\, (t/t_{ch})^{-3/7} \;\; ({\rm for} \, t<t_{ST}) 
\end{equation}
where $u_{ch}=(E_{SN}/M_{ej})^{1/2}$, $t_{ch}= E_{SN}^{-1/2} M_{ej}^{5/6}
\rho_1^{-1/3}$ and $t_{ST}= 0.732 \,t_{ch}$. We adopt the same values of
$E_{SN}$, $M_{ej}$ and $\rho_1$ used in the two cases above.

For what concern the magnetic field we assume that the amplification mechanism
is at work, converting a fraction of the incoming kinetic energy flux into
magnetic energy density downstream of the shock. The simplest way to write this
relation is:
\begin{equation} \label{eq:B2_t}
 B_2^2(t)/(8\pi)= \alpha_B \rho_1 u_1^2(t) \,.
\end{equation}
We always assume $B_2= 4 B_1$. The parameter $\alpha_B$ hides all the
complex physics of magnetic amplification and particle acceleration. For the
sake of completeness we mention here that in the case of resonant streaming
instability $\alpha_B \propto \xi_{cr} v_A/u_1$, where $\xi_{cr}$ is the
efficiency in CRs and $v_A$ is the Alfv\'en velocity, while in the case of
resonant amplification $\alpha_B \propto \xi_{cr} u_1/4c$ \cite[]{bell04}.
However such relations cannot be applied in a straightforward way since they
require using a non linear theory. Here we prefer to make the simplest
assumption, taking $\alpha_B$ as a constant. We determine its value from the
same condition previously used to fix the value of the constant magnetic field:
namely we assume that for $t=t_{ST}$ the energy of protons is $3\cdot 10^{15}$
eV. This condition gives $\alpha_B= 3.25 \cdot 10^{-3}$ and $6.80\cdot 10^{-3}$
for type I/a and core-collapse cases, respectively. 
It is worth noting that in the case of some young SNRs the value of $\alpha_B$
has been estimated from the measurement of both the shock speed and the magnetic
field strength. However we stress that the magnetic field cannot be determined
in a completely model independent way. A possible method requires the
measurement of the spatial thickness of the x-ray filaments: assuming that this
thickness is determined by the rapid synchrotron losses of radiating electrons
we can get the average magnetic field (see Table~\ref{tab:2} and the listed
references and also Table 1 from \cite{caprioli09}). Noticeably the values
estimated for $\alpha_B$ are only a factor $5-10$ larger then the values we use
here.

Now Eq.~(\ref{eq:p2_sol2}) can be used to compute the maximum momentum at
$t=t_{ST}$. For each time-step, $\tau_{\rm ph,k}$, the values of $u_k$ and $B_k$
are computed according to Eqs.~(\ref{eq:u1_t}) and (\ref{eq:B2_t}),
respectively, evaluated at the beginning of the time-step.
In Fig.~\ref{fig:EmaxII} we report the results for the maximum energy for the
case of type I/a (upper panel) and core-collapse SNRs (lower panel). The lines
have the same meanings as in Fig~\ref{fig:EmaxI}. As it is clear from both the
panels, the reduction of the maximum energy is now more pronounced with respect
to the stationary case shown in Fig~\ref{fig:EmaxI}. Even in the scenario of a
core-collapse SNR located in the Galactic bulge, iron nuclei are accelerate only
up to about one half of the maximum theoretical energy. The numerical values of
$\tau_{\rm ion}^{\rm tot}$ and $E_{\max}/E_{\max}^0$ for iron are reported in
Table~\ref{tab:1} in round brackets. One can see that even in the cases where
$\tau_{\rm ion}^{\rm tot}$ does not change with respect to the stationary case,
the maximum energy results smaller.

\begin{figure}
\begin{center}
{\includegraphics[angle=0,width=1.0\linewidth]{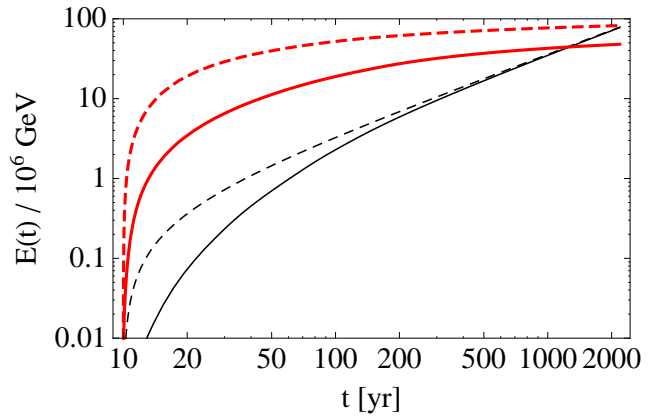}}
\caption{Behavior of maximum energy of Fe ions as a function of the time in
the case of core-collapse SNR located in the Galactic bulge. Two different
scenarios are shown: ({\it thin lines}) acceleration at a constant rate
($B_1=\rm const$ and $u_1=\rm const$); ({\it thick lines}) acceleration
including the time evolution of the remnant ($B_1= B_1(t)$ and $u_1= u_1(t)$).
The dashed lines represent the case of completely ionized ions, while the solid
lines include the effect of photo-ionization.}
\label{fig:EmaxII_Fe}
\end{center}
\end{figure}

The reason why ionization is less effective when the evolution is taken
into account is a consequence of the fact that acceleration occurs mainly during
the first stage of the SNR expansion. In fact both the shock speed and the
magnetic field strength are larger at smaller times. On the other hand the
effective ions' charge is small during the very initial phase of the expansion,
hence the acceleration rate is smaller than its maximum possible value.
In order to clarify this point in Fig.~\ref{fig:EmaxII_Fe} we compare the
energy achieved by iron nuclei as a function of the time, for the case of core
collapse SNR located in the Galactic bulge, with and without the evolution.
The dashed lines shown the case of completely ionized nuclei, while the solid
lines take into account the ionization process. When the acceleration rate is
constant (thin lines) the bulk of the acceleration occurs close to $t_{ST}$,
when the ions are almost completely ionized, hence $E_{\max}$ and $E_{\max}^0$
are practically the same. On the other hand when the evolution is taken into
account (thick lines) the acceleration mainly occurs during the first 200 yr
when the ionization is still not complete and for $t=t_{ST}$ we have
$E_{\max}/E_{\max}^0= 0.58$. Besides the numerical value, this exercise shows
that the inclusion of the time evolution and non linear effects can enhance the
reduction of the maximum energy due to the ionization process.

Some comments are in order. The most relevant consequence of the ionization
mechanism concerns the shape of the CR spectrum in the {\it knee} region.
As we have already discussed, the relation $E_{\max,N} \propto Z_N$ is needed in
order to fit the data. Even a small deviation of the cutoff energy from the
direct proportionality can affect the slope. Let us assume that the measured
slope, $s=3.1$, indeed results from this proportionality relation.
We have $E_{\max, Fe}= 26 E_{\max,p}$. Now if the maximum energy of iron
decreases to 80\% of its maximum value, the slope changes to $\sim 3.33$, while
for a decrease of 50\% the slope becomes $\sim 3.94$.

According to our results we can say that in the context of the SNR paradigm, the
primary sources of CRs above the {\it knee} energy are most probably the
core-collapse SNRs with $M_{ej}\gg 1 M_{\odot}$. Type I/a SNRs seem unable to
accelerate ions up to an energy $Z_N$ times the proton energy, due to their
small Sedov-Taylor age. This consideration is strengthened by the fact that the
CR flux observed at the Earth is mostly due to the SNRs located in the solar
neighborhood, rather than those located in the bulge. In fact the escaping
length from the Galaxy is determined by the thickness of the Galactic halo which
is $\sim 3-5$ kpc, while the Galactic bulge is at 8 kpc from us.

A second comment concerns the acceleration of elements beyond the iron group.
Even if the contribution of such ultra-heavy elements to the CR spectrum is
totally negligible at low energies, at higher energies it can hardly be
measured. In principle the contribution of ultra-heavy elements
could be significant in the 100 PeV regime, which is the energy region where the
transition between Galactic and extragalactic CRs occurs. Indeed several authors
pointed out that a new component is needed to fit this transition region, beyond
the elements up to iron accelerated in ``standard'' SNRs (see e.g. the
discussion in \cite{caprioli10}). As inferred by \cite{horandel03} stable
elements heavier than iron can significantly contribute to the CR spectrum in
the 100 PeV regime if one assume that their maximum energy scales like $Z_N$.
This assumption is especially appealing also for a second reason: in principle
it could explain the presence of the {\it second knee} in the CR spectrum
\cite[]{horandel03}; in fact the ratio $E_{2^{nd}knee} / E_{knee}
\simeq 90$ is very close to the nuclear charge of the last stable nucleus,
uranium, which should have $E_{\max,U}= 92 E_{\max,p}= 414$ PeV.
On the other hand  the ionization mechanism discussed here provides a strong
constraint on the role of ultra-heavy elements. If the acceleration of elements
heavier than Fe occurs at SNRs like those considered above, it is easy to show
that they cannot achieve energies much larger than the Fe itself, because the
total ionization time increases rapidly with the nuclear charge. A contribution
from ultra-heavy elements is maybe possible only if they are accelerated in very
massive SNRs, with $M_{ej}\gg 1 M_{\odot}$. A second possibility that would be
interesting to investigate is the acceleration during the very initial stage of
the remnant, when the expansion occurs into the progenitor's wind. In this case
the high density of the wind could strongly reduce the ionization time thanks to
the Coulomb collisions and nuclei could be totally stripped in a short time.

\section{Solution of the transport equation}  \label{sec:spectra}

In this section we present the solution of the stationary transport equation
for ions and electron accelerated at shocks when the ionization process is taken
into account. We first get the solution for the distribution function of
ions which can be used, in turn, as the source term responsible for the
injection of electrons. We will consider only the test-particle solution,
neglecting all the non linear effects coming from the back reaction of the
accelerated particles onto the shock dynamics.

\subsection{Spectrum of ions}

In order to compute the distribution function of a single chemical specie we
consider separately each ionization state. Let $f_N^Z(x, p)$ be the distribution
function of the specie $N$ with effective charge $Z$. The equation that describe
the diffusive transport of ions in one dimension, including the ionization
process reads:
\begin{equation} \label{eq:trans}
 u \frac{\partial f_N^Z}{\partial x} = D(p) \frac{\partial^{2}
 f_N^{Z}}{\partial x^{2}} + 
 \frac{1}{3}\frac{du}{dx}p\frac{\partial f_N^Z}{\partial p} 
 -\frac{f_N^Z}{\tau_N^Z(p)}  + Q_N^Z(x,p) \,,
\end{equation}
where $u(x)$ is the fluid velocity and $D(p)$ is the diffusion coefficient,
which is a function of $Z$ and it is assumed constant in space. $\tau_N^Z(p)$ is
the ionization time for the losses of one electron, namely for the process $N^Z
\rightarrow N^{Z+1}+e^-$. We neglect multiple ionization processes. In order to
solve Eq.~(\ref{eq:trans}) we need to specify the injection term $Q_N^Z(x,p)$.
We assume that ions with the lower degree of ionization, $Z_0$, are injected
only at the shock position ($x=0$) and at a fixed momentum $p_{\rm inj}$. Hence
the source term for $f_N^{Z_0}$ is:
\begin{equation} \label{eq:Q_Z0}
 Q_N^{Z_0}(x,p) =  K_N^{Z_0} \, \delta(x) \, \delta(p-p_{\rm inj})\,.
% K=\frac{n_1 \eta}{4 \pi p_{\rm inj}^3}
\end{equation}
The normalization constant $K_N^{Z_0}$ is determined from the total number of
particles injected per unit time. For the subsequent ionization states, $Z>Z_0$,
the injection comes from the ionization of atoms with charge equal to $Z-1$,
i.e.:
\begin{equation} \label{eq:Q_Z}
 Q_N^{Z+1}(x,p) =  f_N^{Z}(x,p)/\tau_N^{Z}(p) \,.
\end{equation}
Equation~(\ref{eq:trans}) can be solved separately in the upstream and
downstream regions, where the term $du/dx$ vanishes and the equation reduces to
an ordinary differential equation of second order. We label the quantities in
the upstream (downstream) with a subscript 1(2) and we adopt the convention
$x<(>)0$ in the upstream (downstream). We fix the boundary conditions at the
shock position such that $f(x=0,p) \equiv f_0(p)$, where $f_0(p)$ has to be
determined. The boundary condition at infinity is $\partial f/\partial x=0$ for
$x\rightarrow \pm \infty$. For the sake of simplicity we now drop the label
characterizing the ion specie, such that $f_N^Z \equiv f$, and we write the
solution in a compact form as follows:
\begin{eqnarray} \label{eq:ion_sol1}
 f_i(x,p) = f_0(p) e^{\mp\lambda_{i\mp} x} 
 \pm \int_{x}^{\pm\infty} \frac{Q_i(x',p)}{u_i \sqrt{1+\Delta_i}}
       e^{\pm\lambda_{i\pm}(x-x')} dx'  \nn \\ 
 \pm \int_0^x \frac{Q_i(x',p)}{u_i \sqrt{1+\Delta_i}} 
       e^{\mp\lambda_{i\mp}(x-x')} dx' \nn \\
 \mp e^{-\lambda_{i\mp} x} 
      \int_0^{\pm\infty} \frac{Q_i(x',p)}{u_i \sqrt{1+\Delta_i}}
       e^{-\lambda_{i\pm} x'} dx' \,.
\end{eqnarray}
The upper and lower signs refer to the downstream ($i=2$) and upstream ($i=1$)
solutions, respectively. The dimensionless function $\Delta(p)$ is the ratio
between the diffusion time and the ionization time, i.e.  $\Delta_i(p) \equiv
\frac{4D_i(p)}{u_i^2 \tau(p)}$. $\lambda_{i+}$ and $\lambda_{i-}$ are the
inverse of the propagation lengths along the counter-streaming and the streaming
directions with respect to the fluid motion, respectively, i.e.:
\begin{equation} \label{eq:lambda}
 \lambda_{i\pm}^{-1}(p)= \frac{2 D_i(p)}{u_i [\sqrt{1 + \Delta_i(p)} \pm 1]}\,.
\end{equation}
The quantity $\lambda$ takes into account the diffusion, the advection and
the ionization processes. One can see that in the limit where the ionization is
negligible, namely when $\tau \rightarrow \infty$, the streaming propagation
length diverges, while the counter-streaming one reduces to $D/u$. 

The three integral terms which contribute to the distribution function in  
Eq.~(\ref{eq:ion_sol1}) have a clear physical interpretation. Let us consider
the solution in the upstream: in this case the first integral represents
the contribution due to particles advected with the fluid from $-\infty$, the
second accounts for the particles diffusing in the counter-streaming direction,
while the last integral takes into account the flux escaping across the shock
surface. For the downstream solution the meanings of the first end second
integrals are reversed, while the third one keeps the same meaning. This
interpretation becomes clear if one take the limit $\tau \rightarrow \infty$.

From Eq.~(\ref{eq:ion_sol1}) we can easily recover the standard solution where
the ionization is not taken into account, assuming that the injection occurs
only at the shock position and using $\Delta_i=0$: in this case the integral
terms vanish, and we find $f_1(x,p)=f_0(p) \exp[-u_1 x/D_1]$ and $f_2(x,p)=
f_0(p)$.

In order to get the complete solution we need to compute the boundary
distribution function at the shock position, $f_0(p)$. We follow a standard
technique which consists in integrating Eq.~(\ref{eq:trans}) around the shock
discontinuity \cite[]{Blasi02}. We got the following differential equation for
$f_0(p)$:
\begin{equation} \label{eq:f0}
 \frac{u_1-u_2}{3} p \frac{\partial f_0}{\partial p}
 = \left( D_2 \frac{\partial f_2}{\partial x} \right)_{0^+}
 + \left( D_1 \frac{\partial f_1}{\partial x} \right)_{0^-} \,,
\end{equation}
where $0^+$ and $0^-$ indicate the positions immediately downstream and upstream
of the shock. The quantities $D \partial f/\partial x$ can be easily computed
deriving $f$ with rspect to $x$ from Eq.~(\ref{eq:ion_sol1}). We get:
\begin{equation} \label{eq:sol_D2}
 \left( D_2 \frac{\partial f_2}{\partial x} \right)_{0^+} = 
  f_0 \frac{u_2}{2} \left( 1+\sqrt{1+\Delta_2} \right)
  + \int_0^{\infty} Q_2 e^{-\lambda_{2+} x'} dx'
\end{equation}
and
\begin{equation} \label{eq:sol_D1}
 \left( D_1 \frac{\partial f_1}{\partial x} \right)_{0^-} = 
  f_0 \frac{u_1}{2} \left( 1-\sqrt{1+\Delta_1} \right)
  - \int_{-\infty}^0 Q_1 e^{\lambda_{1-} x'} dx' \,.
\end{equation}
After a little algebra, the solution of Eq.~(\ref{eq:f0}) can be expressed in
the following form:
\begin{eqnarray} \label{eq:ion_sol_f0}
 f_0(p)= s\,p^{-s} \int_{p_{\rm inj}}^{p} \frac{dp'}{p'} p'^s
 \frac{G(p')}{u_1} 
 \exp \left\{-s \int_{p'}^{p} \frac{dp''}{p''} h(p'')
        \right\} \,,
\end{eqnarray}
where the usual definition of the power law index is $s=3u_1/(u_1-u_2)$. The
injection term $G$ is the sum of the upstream and downstream contributions,
\begin{equation} \label{eq:G}
 G(p)= \int_{-\infty}^{0}  Q_1 \, e^{\lambda_{1-} x'}dx'
       + \int_{0}^{\infty} Q_2 \, e^{-\lambda_{2+} x'}dx'
\end{equation}
while the function $h$ in the exponential is: 
\begin{equation} \label{eq:h}
 h(p)= \frac{1}{2}(\sqrt{1+\Delta_1(p)}-1) + \frac{1}{2r}
       (\sqrt{1+\Delta_2(p)}-1) \,.
\end{equation}
The solution in Eq.~(\ref{eq:ion_sol_f0}) shows the typical power law behavior
$\propto p^{-s}$ in the region of momentum where neither injection nor 
ionization are important, in fact when $\Delta(p)\ll1$ then $h(p)$ vanishes.
On the other hand when the ionization length becomes comparable with the
diffusion length, i.e. $\Delta(p)\simeq 1$, the function $h(p)$ produces an
exponential cutoff in the distribution function.

\begin{figure}
\begin{center}
{\includegraphics[angle=0,width=1.0\linewidth]{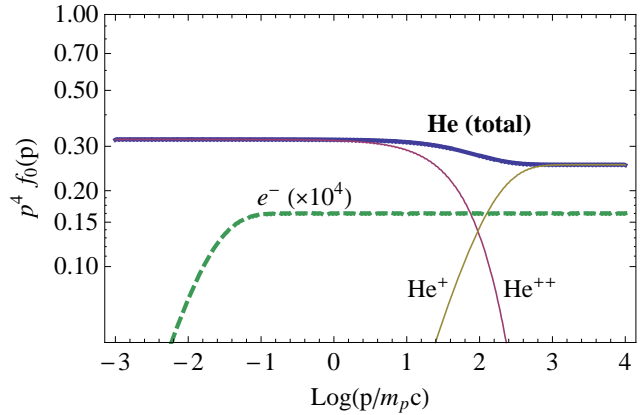}}
\caption{Distribution function of Helium at the shock position, $f_{\rm
He,0}(p)$. The thin solid lines represent the distributions of He$^+$ and
He$^{++}$, while the thick line is the sum of both the contributions. The dotted
lines represent the distribution of electrons (multiplied by $10^4$) ejected in
the process $\rm He^+ \rightarrow He^{++} + e^-$, and subsequently accelerated.}
\label{fig:He}
\end{center}
\end{figure}

\begin{figure}
\begin{center}
{\includegraphics[angle=0,width=1.0\linewidth]{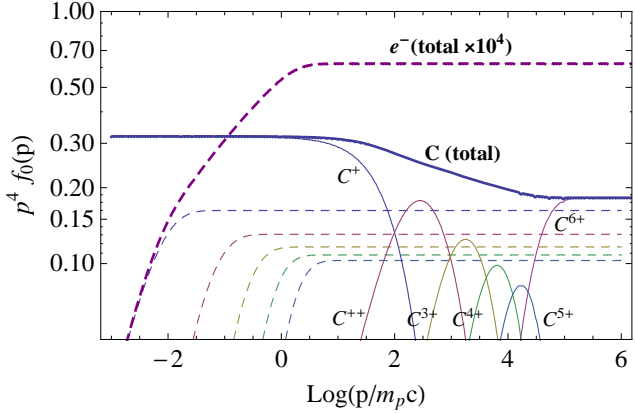}}
\caption{Distribution function of Carbon in all the ionization state from
C$^+$ up to C$^{6+}$. The thick solid line is the sum of all contributions. The
spectra of electrons ejected from each ionization process are plotted with thin
dashed lines, while the thick dashed line represent the total sum
($\times 10^4$).}
\label{fig:C}
\end{center}
\end{figure}

Now that we got the formal solution we can show some results. Let us start
considering the acceleration of He. We assume that the initial ionization state
at the moment of injection is He$^+$. The distribution function $f_{\rm He^+}$
is computed with the procedure described above, using the injection term in
Eq.~(\ref{eq:Q_Z0}). We fix $p_{\rm inj}= 10^{-3} m_pc$ and the shock velocity
equal to 3000 km/s. The photo-ionization time can be an arbitrary function of
$p$. We use the approximate expression given in Eq.~(\ref{eq:Ph_time2}).
Eqs.~(\ref{eq:ion_sol1}) and (\ref{eq:ion_sol_f0}) can be integrated
numerically. Once $f_{\rm He^+}$ is known, we can compute $f_{\rm He^{++}}$
using the injection term in Eq.~(\ref{eq:Q_Z}), i.e. with $Q_{\rm He^{++}} =
f_{\rm He^+}/\tau_{\rm He^+}$. The resulting distributions are plotted in
Fig.~\ref{fig:He} with thin solid lines. We note that $f_{\rm He^+}\propto
p^{-4}$ up to $p\simeq 10 m_pc$; above such value the distribution of He$^+$
drops exponentially while the distribution of He$^{++}$ starts rising. 
It is worth noting that the total distribution (plotted with thick solid line)
is slightly steeper than $p^{-4}$ in the transition region. This effects is due
the different probability which He$^+$ and He$^{++}$ have to return at the shock
once they are downstream. In fact an ion with charge $Z$ located at the position
$x$ in the downstream have a probability to return at the shock $\propto
e^{-u_2x/D(Z)}$; after the ionization the charge becomes $Z+1$ and the return
probability is reduced by a factor $e^{-(Z+1)/Z}$. This translates into a net
loss of particles from downstream. This effect of losses is more pronounced
when considering heavier elements. In Fig.~\ref{fig:C} we show the case of
carbon atoms which start the acceleration as singly ionized. The distribution
functions of all ionized state are plotted with thin solid lines, from $C^+$ up
to $C^{6+}$. In the energy region where ionization occurs, $1.5<\gamma<4.5$, the
slope of the total distribution function is $s\sim 4.1$. 

\subsection{Spectrum of electrons}
Now that we know how to compute the distribution function of ions, the solution
for the electron distribution is straightforward. We have to solve the same
transport equation (\ref{eq:trans}) writing down the right expression for the
electron injection, $Q_e$, and dropping the term due to ionization. In the
present calculation we neglect energy losses.

Let us consider a single specie $N$. From each photo-ionization process of the
type $N^Z \rightarrow N^{Z+1}+e^-$, we have the following contribution to the
electron injection:
\begin{eqnarray} \label{eq:Q_el}
 Q_{e}^{N,Z}(x,p) 
   = \int_p^\infty dp' c\, 
       \frac{f_{N}^z(x,p')}{\tau_N^z(p')} \, 
       \delta^{(3)}\left( p - \xi_N p' \right) \nn \\
   = \xi_N^{-3} \frac{f_{N,1}^z(x,p/\xi_N)}{\tau_N^z(p/\xi_N)} 
\end{eqnarray}
The $\delta$-Dirac function in momentum is present because electrons are ejected
with the same Lorentz factor of the parent ions, as we discuss in
\S\ref{sec:time}. Hence the electron momentum is $p_e=p_N m_e/m_N \equiv p_N
\xi_N$. Following the procedure used in the previous section, we can write the
distribution function for electrons at the shock position as follows:
\begin{equation} \label{eq:sol_el0}
 f_{e,0}^{N,Z}(p)= s \, p^{-s} \xi_N^{s-3} \int_{p_{\rm inj}}^{p/\xi_N} 
             \frac{dy}{y} y^s \frac{G_{e}^{N,z}(y)}{u_1}  \,.
\end{equation}
Equation (\ref{eq:sol_el0}) is similar to Eq.~(\ref{eq:ion_sol_f0}) with the
exception of the multiplying factor $\xi_N^{s-3}$ and for the absence of the
exponential cutoff.
We note that the contribution to electron injection from the downstream is
negligible with respect to the upstream. This occurs because electrons produced
in the downstream have a negligible probability to come back to the shock. The
return probability for one electron stripped downstream at a position $x$ is a
factor $e^{-m_N/Z_N m_e}$ smaller that the return probability of the parent ion.
Hence in the injection term we consider only the contribution from upstream,
which reads:
\begin{equation} \label{eq:G_e}
 G_{e}^{N,Z}(p)= \int_{-\infty}^0 Q_e^{N,Z}(x',p) dx'  \,.
\end{equation}
In Figs.~\ref{fig:He} and \ref{fig:C}, besides the distribution of ions, we plot
also the distributions of the electrons due to each single ionization step,
multiplied by $10^4$ (dashed lines).
Comparing Eqs.~(\ref{eq:ion_sol_f0}) and (\ref{eq:sol_el0}) we see that the
ratio between electron and ion distribution functions is:
\begin{equation} \label{eq:K_eN}
 K_{eN} \equiv f_{e,0}^{N,Z}(p)/f_{N,0}^Z(p)= \xi_N I_{N,Z}(p) \,.
\end{equation}
The function $I_{N,Z}(p)$ represents the ratio between the injection integrals
in Eqs.~(\ref{eq:ion_sol_f0}) and (\ref{eq:sol_el0}). It can be computed
numerically and in the asymptotic limit $p\gg p^*$ (where $p^*$ is such that
$\Delta(p^*)=1$) it is $I_{N,Z} \sim Z/(2Z-1)$. This result is not surprising:
it reflects the fact that for each ionization event electrons can be injected
only if they are released upstream hence their number is roughly 1/2 of the
injected ions.

Now we are interested in the total number of accelerated electrons that can be
produced via the ionization process. In the literature the number of electrons
is usually compared with that of accelerated protons: DSA operates in the same
way for both kinds of particles, hence a proportionality relation between their
distribution functions is usually assumed, i.e. $f_e(p) = K_{ep} f_p(p)$.
From the observational point of view, the value of $K_{ep}$ in the source
strongly depends on the assumption for the magnetic field strength in the region
where electrons radiate and, in the context of the DSA theory, it can be
determined for those SNRs where both non thermal X-ray and TeV radiation are
observed. Two possible scenarios have been proposed. In the first one electrons
produce both the X-ray and the TeV components via the synchrotron emission and
the inverse Compton effect, respectively; this scenario requires a downstream
magnetic field around $20\mu$G, and $K_{ep}\sim 10^{-2}-10^{-3}$. The second
scenario assumes that the number of accelerated protons is large enough to
explain the TeV emission as due to the decay of neutral pions produced in
hadronic collisions. In this case the DSA requires a magnetic field strength of
few hundreds $\mu$G and $K_{ep} \sim 10^{-4}-10^{-5}$. Such a large magnetic
field is consistently predicted by the theory as a result of the magnetic
amplification mechanisms which operate when a strong CR current is present.
In Table~\ref{tab:2} we report some examples of young SNRs where the nonlinear
theory of acceleration has been applied to fit the multi-wavelength spectrum.
The forth and fifth columns report the estimated valued of the downstream
magnetic field strength and the corresponding electron/proton ratio.

\begin{figure}
\begin{center}
{\includegraphics[angle=0,width=1.0\linewidth]{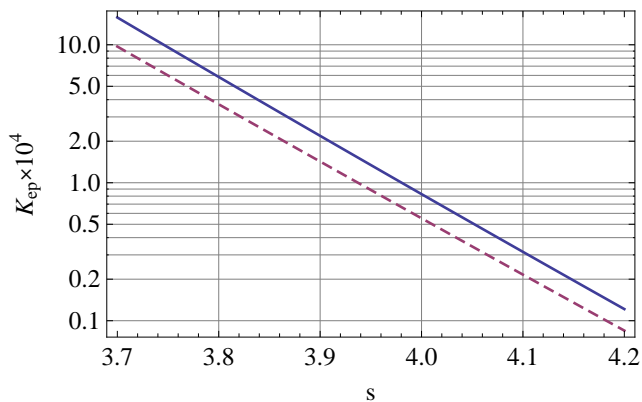}}
\caption{Value of $K_{ep}$ computed from Eq.~(\ref{eq:K_ep}) as a function of
the slope $s$ and for $\delta=0.6$ (solid line) and $\delta=0.3$ (dashed line).}
\label{fig:Kep}
\end{center}
\end{figure}

In order to give an approximated estimate for $K_{ep}$ we need to multiply
Eq.~(\ref{eq:K_eN}) by the total number of ejected electrons, i.e.
$(Z_N-Z_{N,0})$, and sum over all atomic species present in the accelerator,
i.e. $K_{ep}\simeq \sum_N K_{Np}\,(Z_N-Z_{N,\rm eff}) \,K_{eN}$. $K_{Np}$
are the abundances of ions measured at the source in the range of energy where
the ionization occurs. Even if the values of $K_{Np}$ are widely unknown, we
can estimate them using the abundances measured at Earth and adding a correction
factor to compensate for propagation effects in the Galaxy, namely the fact that
particles with different $Z$ diffuse in a different way. The diffusion time in
the Galaxy is usually assumed to be $\tau_{\rm diff}\propto (p/Z)^{-\delta}$,
with $\delta\approx 0.3-0.6$ (see \cite{Blasi07} for a review on recent CR
experiments). If $K_{Np,0}$ is the ion/proton ratio measured at Earth, than the
same quantity measured at the source is $K_{Np}= K_{Np,0}Z_N^{-\delta}$. Hence
the final expression for the electron/proton ratio at the source is: 
\begin{equation} \label{eq:K_ep}
 K_{ep}= \sum_N K_{Np,0}\, Z_N^{-\delta}\,(Z_N-Z_{N,0})
  \frac{Z_N}{2Z_N-1} \left(\frac{m_e}{m_N} \right)^{s-3} \,.
\end{equation}
In Fig.~\ref{fig:Kep} we report the value of $K_{ep}$ computed according to
Eq.~(\ref{eq:K_ep}) as a function of the spectral slope $s$ and for $\delta=0.3$
and 0.6. For $K_{Np,0}$ we use the abundances of nuclei in the CR spectrum
measured at 1 TeV \cite[]{wiebel}. Moreover we determine the initial charge of
different species, $Z_{N,0}$, adopting the thermal equilibrium in a plasma with
a temperature $T\sim 10^5$. Remarkably using the slope predicted by linear
theory for strong shock, $s=4$, we have $K_{ep} \sim 10^{-4}$: this number gives
the right order of magnitude required to explain the values of $K_{ep}$ reported
in Table~\ref{tab:2}.
We stress here that in order to provide a better estimate of $K_{ep}$ for a
specific SNR we should know the local chemical composition and have a realistic
model for the injection of heavy elements into the DSA, which can provide the
right value of $Z_{N,0}$. 

\begin{table}
\begin{center}
\begin{tabular}{cccccc}
\hline
SNR & Age (yr) & $d({\rm kpc})$ & $B_2 ({\rm \mu G})$ & $K_{ep}/10^{-4}$ 
& ref. \\
\hline
\hline
Kepler     & 400     & 3.4--7.0 & 440-500   & 0.7--2.8 & $^{(a)}$  \\
Tycho      & 430     & 3.1--4.5 & 368-420   & 4.2--15  & $^{(b)}$  \\
SN 1006    & 1000    & 1.8      & 90        & 1.3      & $^{(c)}$  \\
   ''      &  ''     & 2.2      & 120       & 4.1      & $^{(d)}$  \\
G347.3-0.5 & 1600(?) & 1.0      & 100       & 0.6      & $^{(e)}$  \\
   ''      &  ''     & 1.0      & 130       & 1.0      & $^{(f)}$  \\
\hline
\end{tabular}
\caption{Estimated value for the magnetic field and $K_{ep}$
for some young supernova remnants. Results are from:
$^{(a)}$Berezhko et al. (2006),
$^{(b)}$V\"olk et al. (2008),
$^{(c)}$Morlino et al. (2010),
$^{(d)}$Berezhko et al. (2009),
$^{(e)}$Morlino et al. (2009),
$^{(f)}$Berezhko \& V\"olk (2006).
} \label{tab:2}
\end{center}
\end{table}

Equation~(\ref{eq:K_ep}) has been obtained using the test particle
approach. A better estimate requires the use of nonlinear theory which is,
indeed, much more complicated to develop. Nevertheless it is easy to realize
that the main effect that nonlinearity can produce on $K_{ep}$ is due to the
different prediction of the slope $s$. In fact, while the test particle approach
gives the universal slope $s=4$, the nonlinear theory predict concave spectra
with $s=s(p)$. The basic formulation of the nonlinear theory usually leads to
$s>4$ for momenta lower that $\sim 10 Z^{-1}$ GeV/c and $s<4$ for larger momenta
\cite[]{amato05}. As discussed in \S\ref{sec:time}, ionization typically
occurs for $\gamma \gtrsim 10$, where the slope is $s<4$. Hence we can conclude
that the nonlinear effects tend to increase the value of $K_{ep}$. For instance,
from Fig.~\ref{fig:Kep} we see that for $s=3.7$, $K_{ep}$ rises up to $\sim
10^{-3}$.

It is worth noting that nonlinear effects can also result in the opposite
situation with $K_{ep}<10^{-4}$, as we will show below. A key ingredient of
the DSA is the speed of the magnetic turbulence which is responsible for 
particles scattering. The speed of the scattering centers contributes to the
determination of the effective compression ratio felt by particles which, in
turns, determines the spectral slope. The typical speed of the turbulence is of
the order of Alfv\'en speed which, when is computed in the background magnetic
field, is of the order of few tens km/s and it is negligible with respect to the
shock speed. On the other hand if one assume that turbulence moves with the
Alfv\'en speed computed in the amplified magnetic field, as shown by
\cite{caprioli10}, it cannot be neglected and the resulting slope can be
$>4$. If this happens the value of $K_{ep}$ drops and the number of electrons
produced via ionization is maybe insufficient in order to explain the
synchrotron emission.

A final comment concerns the mechanism of electrons injection in core-collapse
SNRs. The value of $K_{ep}$ shown in Fig.~\ref{fig:Kep} has been computed using
the ionization level of chemical elements in a plasma with $T\sim 10^5$. We
recall that the type I/a SNRs expand in the warm ISM, where the typical
temperature is around $10^4$ K hence we expect the injection of electrons to be
relevant. On the other hand core-collapse SNRs expands into a diluted bubble
whose temperature reach $10^6$ K \cite[]{higdon}, hence the degree of ionization
of atoms is higher and the number of electrons available for the injection is
smaller (see e.g. \cite{porquet}). Using $T = 10^6$ we estimate that $K_{ep}$ is
a factor $\sim 4$ smaller than the values shown in Fig.~\ref{fig:Kep}. On the
other hand in those bubbles the metallicity is estimated to be greater than the
ISM mean value \cite[]{higdon}. Hence, even for core-collapse SNRs the electron
injection through ionization could play a relevant role.

\section{Discussion and conclusions}  \label{sec:conclusion}
In this work we include for the first time the ionization process of heavy ions
in the theory of DSA at SNR shocks. We showed that, for the typical environments
where SNRs propagate, the photo-ionization due to the ISRF dominates on the
collisional ionization and produces two important effects: 1) the reduction of
the maximum energy achieved by ions and 2) the production of a relativistic
population of electrons. 

(1) The first effect is especially important for the interpretation of the knee
structure in the all particle spectrum of CRs. The change of slope on the two
sides of the knee is generally interpreted as due to the superposition of
spectra of chemicals with different nuclear charges combined with their
abundances and convolved with the rigidity dependent diffusion in the Galaxy. 
An important requirement for a good fit to the knee is that the cutoff
energy of each specie, from hydrogen up to iron, should be proportional to the
nuclear charge, $Z_N$.
 
It is generally assumed that DSA can accelerate ions up to a maximum energy
proportional to $Z_N$. This assumption is valid only if the ionization time
needed to strip all electrons from atomic orbitals, $\tau_{\rm ion}^{\rm tot}$,
is much shorter than the acceleration time. We assume that the maximum possible
acceleration time is equal to the Sedov-Taylor time, $t_{ST}$, corresponding to
the end of the free expansion phase. Using a simple steady-state approach we
showed that, especially for very massive ions, $\tau_{\rm ion}^{\rm tot}$ can be
comparable to or even larger than $t_{ST}$, depending on the type of the
remnant and on its location in the Galaxy. In fact the ISRF responsible for the
photo-ionization decreases for increasing distance from the Galactic Center. We
analyze two different cases which are representative of a type I/a and a
core-collapse SNRs.
Our main results are the following. SNRs of type I/a are generally unable to
accelerate ions according to $E_{\max,N} \propto Z_N$ because $t_{ST}$ does not
exceed the typical value of 500 yr. The only exception are the remnants located
in the Galactic bulge where the photon field is strong enough to reduce the
photo-ionization time to a value much shorter than $t_{ST}$. The situation for
core-collapse SNRs is different because their typical Sedov time is $\gtrsim
2000$ yr hence the ionization time for chemicals up to iron can be neglected and
the maximum energy is indeed proportional to the nuclear charge, with the
exception of those remnant located very far away from the Galactic Center where
the ISRF is very low.

Previous conclusions are based on the assumption that during the free expansion
phase both the magnetic field and the shock velocity remain constant. This is
indeed a poor approximation. In fact during the free expansion phase the shock
speed can vary by a factor of few (depending on the velocity profile of the
ejecta and on the density profile of the external medium). Now for DSA
the acceleration rate is proportional to $u_{sh}^2 \delta B(u_{sh})$, where
$\delta B(u_{sh})$ is the turbulent magnetic field generated by the CR induced
instabilities (resonant or non-resonant), which is an increasing function of the
shock velocity. Hence the acceleration rate can significantly change during the
free expansion phase even if $u_{sh}$ varies only by a factor of few. Always in
the framework of steady-state approximation, we develop a simple toy model able
to include the magnetic field amplification and the time evolution of the
remnant. Using this toy model we showed that the magnetic amplification and time
evolution can significantly reduce the acceleration time with respect to
$t_{ST}$ and, as a consequence, the maximum energy achieved by heavy ions
becomes smaller than the previous estimate: the relation $E_{\max,N} \propto
Z_N$ can be achieved up to iron nuclei only by SNRs with a Sedov time $\gg
10^3$ yr which means that the mass ejecta should be $\gg 1\, M_{\odot}$.

Clearly the same mechanism also applies to ions heavier then Fe. Their
maximum energy cannot be much larger than that achieved by Fe itself unless the
acceleration occurs in very massive SNRs. Hence the ionization mechanism put
severe limitation to the possibility that ultra heavier ions can contribute to
the CR spectrum in the transition region between Galactic and extragalactic
component.

(2) The second effect concerning the production of relativistic electrons was
already put forward in \cite{mor09}. The ionization mechanism provides a source
for the injection of relativistic electrons into the DSA mechanism. We
investigate the possibility that those electrons can be responsible for the
synchrotron radiation observed from young SNRs. In order to estimate the total
number of injected electrons we use a semi-analytical technique to solve the
steady-state transport equation which describe the electron distribution
function in the shock region. Summing the contributions coming from the
ionization of all chemicals we showed that the ratio between accelerated
electrons and protons is $K_{ep} \sim 10^{-4}$. This value is especially
appealing because corresponds to the right order of magnitude required in order
to explain the synchrotron emission if the magnetic field is amplified up to few
hundreds $\mu$G. Interestingly such magnetic field amplification is consistently
predicted by non linear acceleration theory as a consequence of the back
reaction of accelerated particles onto the shock dynamics. 

Several effects can modify our prediction for the number of electrons. The
estimate $K_{ep} \sim 10^{-4}$ is obtained using test particle approximation
and adopting the abundances of CR spectrum observed at the Earth corrected for
the effect of propagation in the Galaxy. 
Even if we did not develop a fully nonlinear approach, we showed how nonlinear
effects can substantially modify the electron/proton ratio. Indeed the value of
$K_{ep}$ strongly depends on the spectral slope as one can see from
Eq.~(\ref{eq:K_ep}). Nonlinear DSA usually predicts steeper spectra than the
test particle result, $s=4$, and this translates into an enhancement of
$K_{ep}$.
On the other hand also the opposite situation can be realized. Indeed some
authors pointed out that, if the magnetic field amplification occurs, the
velocity of the magnetic turbulence with respect to the plasma can be much
larger than the Alfv\'en speed corresponding to the background magnetic field .
If this occurs the spectral slope of the accelerated particles be appreciably
softer than 4 and the electron/proton ratio would drop to a value which is
insufficient to explain the observed synchrotron radiation. 

In order to get a correct determination of $K_{ep}$ for a single SNR, other
effects should be taken into account: $a$) the chemical composition of the
environment where the SNR expands, $b$) the initial ionization state of each
chemicals and $c$) a realistic model for the ions injection. A further
complication directly related to the previous points is the fact that many heavy
elements are condensed in solid dust grains both in the ISM as well as in the
stellar wind. It has been pointed out that the injection of refractory elements
into the DSA can be dominated by the sputtering of these dust grains rather than
by the injection of single atoms from the thermal bath \cite[]{ellison97}. In
this picture a correct computation of the number of injected ions and their
initial ionization state is very challenging.

As a finally remark we want to stress that the electron/proton ratio in a single
source, what we call $K_{ep}$, is different from the same ratio measured from
the CR spectrum at Earth, which is $5 \cdot 10^{-3}$ at 1 TeV \cite[]{Blasi07}.
Sometimes these quantities are assumed to be the same. Conversely they could be
different because the latter is the sum of the contribution coming from all
sources integrated during the source age, and also reflect transport to Earth
and losses in transport. Especially relevant is the fact that other sources like
pulsar wind nebulae can significantly contribute to the electron flux measured
at Earth, but not to the proton flux. Moreover the value of $K_{ep}$ in a single
source can vary during the source age. Different mechanisms of electron
injection, other than the ionization, can be relevant in different phases of the
remnant.

\section*{Acknowledgments}
I am grateful to P. Blasi, E. Amato and D. Caprioli for useful and continuous
discussions and exciting collaboration. I also wish to thank the {\it Kavli
Institute for Theoretical Physics} in Santa Barbara where part of this work was
done during the program {\it Particle Acceleration in Astrophysical Plasmas},
July 26-October 3, 2009.
This research was founded through the contract ASI-INAF I/088/06/0 (grant
TH-037).

\end{document}